\documentclass[a4paper]{article}
\usepackage{amsfonts}
\usepackage{amssymb}
\usepackage[margin=2cm]{geometry}
\usepackage[dvips]{graphicx}

\usepackage[T1]{fontenc}
\usepackage[latin2]{inputenc}

\usepackage{amsmath}
\usepackage{bbm}

\begin{document}


\title{Isotropic quantum spin channels and additivity questions}

\author{Robert Alicki\\ 
  Institute of Theoretical Physics and Astrophysics, University
of Gda\'nsk, \\ Wita Stwosza 57, PL 80-952 Gda\'nsk, Poland}

\date{\today}
\maketitle

\begin{abstract}
The minimum entropy output is computed for rotationally invariant quantum channels acting on  spin-1/2 and spin-1 systems. For the case of two parallel such channels and initial entangled (singlet) state the entropy of the output is higher then the doubled minimal entropy output of the single channel. This gives a certain moral support to the additivity hypothesis.
Another related simple function of the channel (minimum entropy gain)
is shown to be additive in general.
\end{abstract}

\section{Introduction}
For a completely positive map $\Lambda$ acting on density matrices $\rho$ of a certain quantum system (quantum channel)
we may define the number ({\em minimal entropy output})
\begin{equation}
S[\Lambda] = \inf_{\rho}S(\Lambda(\rho)) = \inf_{|\phi>}S(\Lambda(|\phi><\phi|))
\label{defmin}
\end{equation}
where $S(\rho)= -{\rm tr}(\rho\log\rho)$ is the von Neumann entropy of the density matrix $\rho$ and the equality of infima in (\ref{defmin}) follows from its concavity .

For a fixed $\Lambda$ we can define the following function of density matrices which can be called
{\em entropy gain}
\begin{equation}
F_{\Lambda}(\rho)= S(\Lambda(\rho))- S(\rho)\ .
\label{defF}
\end{equation}
Computing its infimum over either pure or general mixed states we obtain two functions of the completely positive maps, minimal entropy output and {\em minimal entropy gain} respectively
\begin{equation}
S[\Lambda] =  \inf_{|\phi>}F_{\Lambda}(|\phi><\phi|)
\label{defS}
\end{equation}
\begin{equation}
G[\Lambda] = \inf_{\rho}F_{\Lambda}(\rho)\ .
\label{defG}
\end{equation}
The function $S[\Lambda]$ plays an important role in the quantum information theory and particularly its additivity would be relevant for the problems of entanglement characterization and channel capacities \cite{Shor}. Obviously
subadditivity
\begin{equation}
S[\Lambda\otimes\Lambda'] \leq S[\Lambda]+S[\Lambda']
\label{subS}
\end{equation}
holds and the additivity is reached on product states. Therefore, to prove additivity one should show that taking entangled states $\phi_{AA'}$ is not the optimal strategy to minimize $S(\Lambda\otimes\Lambda'(|\phi_{AA'}><\phi_{AA'}|))$. 
To illustate this issue we shall consider the following bistochastic completely positive maps acting on the spin-$s$ systems 
\begin{equation}
\Lambda_s(\rho)= \frac{1}{s(s+1)}\sum_{k=1}^3 S_k\rho S_k
\label{smap}
\end{equation}
where ${\bf S} = (S_1,S_2,S_3)$ are spin operators satisfying appropriate commutation relations and providing an irreducible representation $\{U_s({\bf n}); {\bf n}\in {\bf R}^3, |{\bf n}|\leq 1\}$ of the $O(3)$ (in fact $SU(2)$) group on the Hilbert space of the dimension $2s+1$. The map (\ref{smap}) satisfies rotational invariance relation
\begin{equation}
\Lambda_s\bigl(U_s({\bf n})\rho U_s^{\dagger}({\bf n}))=U_s({\bf n})\Lambda_s(\rho) U_s^{\dagger}({\bf n}) 
\label{inv}
\end{equation}
and has been used in the context of quantum chaos  \cite{AF01}\cite{AMM96} .

On the other hand we shall prove that the function $G[\Lambda]$ is always additive
\begin{equation}
G[\Lambda\otimes\Lambda'] = G[\Lambda] + G[\Lambda']\ .
\label{adG}
\end{equation}
\section {Minimum entropy output}
We compute the minimal entropy output for the isotropic channels (\ref{smap}) in the case of $s = 1/2, 1$.
The results are summarized in the following theorem.

{\bf Theorem 1}
{\em With the notation of above}
\begin{equation}
S[\Lambda_{1/2}] = \log 3 -\frac{2}{3}\log 2
\label{S1/2}
\end{equation}
\begin{equation}
S[\Lambda_1] = \log 2\ .
\label{S1}
\end{equation}
{\bf Proof}

{\em Case $s=1/2$}

Spin operators are given in terms of Pauli matrices , $S_k = \frac{1}{2}\sigma_k$. The direct computation provides the following useful form of the map $\Lambda_{1/2}$
\begin{equation}
\Lambda_{1/2}(\rho) = ({\rm tr}\rho) \frac{1}{3}{\bf 1} + \frac{1}{3}T\rho T^{\dagger}
\label{l1/2}
\end{equation}

where $T$ is a {\em time-reversal} antiunitary map defined by \cite{Sch}
\begin{equation}
T|+1/2> = |-1/2>\ ,\  T|-1/2> = -|+1/2>\ .
\label{F}
\end{equation}
Therefore, all density matrices $\Lambda_{1/2}(|\phi><\phi|)$ possess the same spectrum $(\frac{1}{3},\frac{2}{3})$
what implies the formula (\ref{S1/2}). The same conclusion follows directly from the invariance (\ref{inv}) which in the case $s=1/2$ means the invariance with respect to all unitary transformation and from the direct computation of
$\Lambda_{1/2}(|+1/2><+1/2|)$. 

{\em Case $s=1$}

The spin operators are given explicitly by the matrices
\begin{equation}
S_1 =\frac{1}{\sqrt{2}}
\begin{pmatrix}
0 & 1 & 0\\
1 & 0 & 1 \\
0 & 1 & 0 
\end{pmatrix}, \ 
S_2 =\frac{1}{\sqrt{2}}
\begin{pmatrix}
0 & -i & 0\\
i & 0 & -i \\
0 & i & 0 
\end{pmatrix},\ 
S_3 =
\begin{pmatrix}
1 & 0 & 0\\
0 & 0 & 0 \\
0 & 0 & -1 
\end{pmatrix}\ .                   
\label{spin}
\end{equation}
In the first step of the proof we show that  
\begin{equation}
S[\Lambda_1] \geq \log 2\ .
\label{S2}
\end{equation}
We use the equality \cite{AF01}
\begin{equation}
S(\Lambda(|\phi><\phi|) = S(\Omega)\ ,\ {\rm where}\ \Omega_{kl}= <\phi , X_lX_k\phi>\ .
\label{sig}
\end{equation}
The matrix $\Omega$ is a $3\times 3$ positively defined matrix with trace one. To prove (\ref{S2}) it is enough to show 
that its operator norm satisfies inequality
\begin{equation}
\|\Omega\| \leq \frac{1}{2}\ .
\label{norm}
\end{equation}
We prove (\ref{norm}) by taking an arbitrary normalized vector $\xi$ from ${\bf C}^3$ and estimating the matrix element
\begin{equation}
<\xi ,\Omega \xi> \leq \frac{1}{2}\|Y^{\dagger}Y\|
\label{normY1}
\end{equation}
where
\begin{equation}
Y = \sum_{k=1}^3\xi_kS_k = (\cos\alpha) ({\bf n}\cdot{\bf S})+
i(\sin\alpha)({\bf m}\cdot{\bf S}).
\label{normY2}
\end{equation}
with two normalized vectors ${\bf n,m}\in{\bf R}^3$ and a certain $\alpha\in[0,\pi]$.
Rotating $Y$ by a natural unitary representation of $O(3)$ we can replace $Y$ in (\ref{normY1}) by the matrix $Z$ of the form
\begin{equation}
Z = (\cos\alpha)  S_3+
i(\sin\alpha)\bigl((\cos\beta) S_3 + (\sin\beta) S_1\bigr)\ .
\label{defZ}
\end{equation}
Introducing a new parametrization 
\begin{equation}
a= \cos\alpha\ ,\ b= \sin\alpha\cos\beta\ ,\ c= \frac{1}{\sqrt{2}}\sin\alpha\sin\beta\ ,a^2 + b^2 +2c^2 = 1
\label{normY2}
\end{equation}
and using (\ref{spin}) we obtain the characteristic polynomial of the matrix $ZZ^{\dagger}$
\begin{equation}
W(\lambda) = {\rm det}
\begin{pmatrix}
a^2+b^2+c^2-\lambda & -ic(a+ib)& c^2\\
ic(a-ib) & 2c^2-\lambda & -ic(a-ib) \\
c^2 & ic(a+ib) & a^2+b^2+c^2-\lambda 
\end{pmatrix}
=
{\rm det}
\begin{pmatrix}
1-\lambda & 0 & 1-\lambda\\
... & ... & ... \\
... & ... & ... 
\end{pmatrix}\ .
\label{det}
\end{equation}
The second determinant in (\ref{det}) was obtained by adding the lowest row to the others and from its structure it follows
that $\lambda =1$ is an eigenvalue of $ZZ^{\dagger}$. Because $ZZ^{\dagger}$ is a positive  matrix with trace equal 2 than it follows that its norm $\|ZZ^{\dagger}\|= \|YY^{\dagger}\|=1$. Hence we have proven the inequalities (\ref{norm}) and (\ref{S2}).

In the last step one can compute explicitly the matrix $\Lambda_1(|1><1|)$  what gives $S(\Lambda_1(|1><1|)=\log 2$. Again from the invariance (\ref{inv}) it follows that the minimal entropy $\log 2$ is reached for all {\em $SU(2)$- coherent states} .

\section{Decoherence of entangled states}
To gain some intuition about the behaviour of $S[\Lambda\otimes\Lambda']$ we can compute the entropy $S(\Lambda_s\otimes\Lambda_s(|0;0><0;0|))$ with a singlet state $|0;0>$. Due to (\ref{inv}) and the invariance of a singlet state with respect to rotations we know that $\rho_s^0 = \Lambda_s\otimes\Lambda_s (|0;0><0;0|)$ is also invariant with respect
to $O(3)$ and therefore has the following structure
\begin{equation}
\rho_{1/2}^0 = \sum_{j=0}^{2s} p_j\frac{1}{2j+1}{\bf I}_{2j+1}
\label{mat}
\end{equation}
where ${\bf I}_{2j+1}$ are projectors on the carriers of irreducible representations in the decomposition od the product
$(2s+1)\otimes(2s+1)$ $O(3)$ representations \cite{Sch}. The probabilities $p_j$ can be computed from the formula
\begin{equation}
<j;0|\rho_{1/2}^0|j;0> =  p_j\frac{1}{2j+1}\ ,\ {\rm hence}\ p_j = (2j+1)<j;0|\rho_{1/2}^0|j;0>
\label{prob}
\end{equation}
which leads to the following expression
\begin{equation}
 p_j = \frac{2j+1}{[s(s+1)]^2}\sum_{k,l=1}^3 |<j;0|S_k\otimes S_l|0;0>|^2\ .
\label{prob1}
\end{equation}
Having the values of $p_j$ we can compute using (\ref{mat})
\begin{equation}
S(\Lambda_s\otimes\Lambda_s(|0;0><0;0|))=\sum_{j=0}^{2s}\bigl(p_j\log(2j+1)- p_j\log p_j\bigr) \ .
\label{entsing}
\end{equation}
The results of computations in the cases $s= 1/2 , 1$ are given below.

{\bf Theorem 2}

{\em With the notation of above} 
\begin{equation}
S(\Lambda_{1/2}\otimes\Lambda_{1/2}(|0;0><0;0|))=\frac{5}{3}\log 3 -\frac{2}{3}\log 2 > 2 S[\Lambda_{1/2}] \ .
\label{th21}
\end{equation}
\begin{equation}
S(\Lambda_1\otimes\Lambda_1(|0;0><0;0|))= \log 3 + \frac{4}{3}\log 2 > 2 S[\Lambda_1] \ .
\label{th22}
\end{equation}
{\bf Proof}

{\em Case $s=1/2$}

It is enough to find $p_0$  by direct computation using (\ref{prob1}) with the form of the singlet \cite{Sch}
\begin{equation}
|0;0>= \frac{1}{\sqrt{2}}\bigl(|+1/2>|-1/2> - |-1/2>|+1/2>\bigr) \ .
\label{sin12}
\end{equation}
One obtains $p_0 = \frac{1}{3} , p_1 =\frac{2}{3}$.
\par
{\em Case $s=1$}

One has to compute $p_0$ and $p_1$ using again (\ref{prob}) with the relevant states \cite{Sch}
\begin{equation}
|0;0>= \frac{1}{\sqrt{3}}\bigl(|+1>|-1> - |0>|0> + |-1/2>|+1/2>\bigr) \ .
\label{sin1}
\end{equation}
\begin{equation}
|1;0>= \frac{1}{\sqrt{2}}\bigl(|+1>|-1> - |-1>|+1>\bigr) \ .
\label{tri}
\end{equation}
The result is $p_0 = \frac{1}{3} , p_1 =\frac{1}{4}, p_2 = \frac{5}{12}$.
\section{Minimal entropy gain}

The properties of the function $G[\Lambda]$ defined by (\ref{defG}) are summarized in the following theorem.

{\bf Theorem 3}

{\em 1) For a system with $d$-dimensional Hilbert space
\begin{equation}
-\log d\leq G[\Lambda]\leq 0\ ,
\label{th31}
\end{equation}
\par
and equal to zero for bistochastic $\Lambda$.

2) $G[\Lambda]$ is additive
\begin{equation}
G[\Lambda\otimes\Lambda'] = G[\Lambda] + G[\Lambda']\ .
\label{th32}
\end{equation}}
{\bf Proof} 

The statement 1) follows directly from the definition (\ref{defG}) taking into account that for a finite 
dimension there exists always an invariant state ${\bar\rho}= \Lambda({\bar\rho})$ and the fact that for a bistochastic map $S(\Lambda(\rho))\geq S(\rho)$.

To prove the statement 2) one can notice that $F_{\Lambda}(\rho)$ is additive for product states and hence
\begin{equation}
G[\Lambda\otimes\Lambda'] \leq G[\Lambda] + G[\Lambda']\ .
\label{p31}
\end{equation}
Applying monotonicity of the relative entropy $S(\rho | \sigma) = {\rm tr}(\rho\log\rho - \rho\log\sigma)$ with respect to completely positive maps \cite{Lin} to the state $\rho_{AA'}$
of the composed system with marginal states $\rho_A$ and $\rho_{A'}$
\begin{equation}
S\bigl(\Lambda\otimes\Lambda'(\rho_{AA'})|\Lambda(\rho_A)\otimes\Lambda'(\rho_{A'})\bigr) \leq
S\bigl(\rho_{AA'}|\rho_A\otimes\rho_{A'}\bigr)
\label{p32}
\end{equation}
one obtains
\begin{equation}
F_{\Lambda\otimes\Lambda'}(\rho_{AA'})\geq F_{\Lambda}(\rho_A) + F_{\Lambda'}(\rho_{A'})\ .
\label{p31}
\end{equation}
This implies $G[\Lambda\otimes\Lambda'] \geq G[\Lambda] + G[\Lambda']$ and completes the proof.

\section{Concluding remarks}

We have analized the class of quantum channels corresponding to the isotropic environment acting on the spin system.
Minimal entropy output has been computed for the cases of spin-1/2 and spin-1. For the doubled channels it is shown
that the initial entanglement of the singlet state increases decoherence and leads to a higher entropy output than 
for the optimal product states. These results provide a certain support for the additivity hypothesis. In addition a new
additive function of the channel - minimal entropy gain - is defined and discussed. 

It is a pleasure to thank  Micha\l\ and Ryszard Horodecki for numerous discussions.   Financial support
by the Polish Ministry of Scientific Research and Information Technology- grant PBZ-MIN-008/P03/2003 and EC grant RESQ IST-2001-37559  is gratefully acknowledged.

\end{document}